\documentclass[12pt]{article}

\renewcommand{\d}{{\rm d}}
\newcommand{\dotOm}{\dot{\Omega}}

\newcommand{\dagOm}{\Omega^\dagger}
\newcommand{\dagu}{u^\dagger}
\newcommand{\dotu}{\dot{u}}
\newcommand{\vf}{\mbox{\boldmath$f$}}
\newcommand{\vn}{\mbox{\boldmath$n$}}

\newcommand{\vb}{\mbox{\boldmath$b$}}
\newcommand{\vl}{\mbox{\boldmath$l$}}
\newcommand{\ve}{\mbox{\boldmath$e$}}
\newcommand{\vg}{\mbox{\boldmath$g$}}
\newcommand{\vnzero}{\vn^{\scriptscriptstyle (0)}}

\newcommand{\dotvn}{\dot{\vn}}
\newcommand{\dotve}{\dot{\ve}}

\newcommand{\prvn}{\vn^\prime}
\newcommand{\prve}{\ve^\prime}
\newcommand{\prdotvn}{\dotvn^\prime}

\newcommand{\pprvn}{\vn^{\prime\prime}}
\newcommand{\ppprvn}{\vn^{\prime\prime\prime}}
\newcommand{\pprve}{\ve^{\prime\prime}}
\newcommand{\pprdotvn}{\dotvn^{\prime\prime}}
\newcommand{\pprdotve}{\dotve^{\prime\prime}}

\newcommand{\llangle}{\sqrt{1 - (\vl_1 \cdot \vl_2)^2}}
\newcommand{\3}{{31^+}}
\newcommand{\0}{{02}}
\newcommand{\nllangle}{\sqrt{1 - (\vl_\3 \cdot \vl_\0)^2}}

\newcommand{\tvn}{\tilde{\vn}}
\newcommand{\dottvn}{\dot{\tvn}}

\renewcommand{\cosh}{{\rm ch}}
\renewcommand{\sinh}{{\rm sh}}

\begin{document}

\title{Spectrum of area in the Faddeev formulation of gravity}
\author{V.M. Khatsymovsky \\
 {\em Budker Institute of Nuclear Physics} \\ {\em of Siberian Branch Russian Academy of Sciences} \\ {\em
 Novosibirsk,
 630090,
 Russia}
\\ {\em E-mail address: khatsym@inp.nsk.su}}
\date{}
\maketitle
\begin{abstract}
Faddeev formulation of general relativity (GR) is considered where the metric is composed of ten vector fields or a ten-dimensional tetrad. Upon partial use of the field equations, this theory results in the usual GR.

Earlier we have proposed first-order representation of the minisuperspace model for the Faddeev formulation where the tetrad fields are piecewise constant on the polytopes like 4-simplices or, say, cuboids into which ${\rm I \hspace{-3pt} R}^4$ can be decomposed, an analogue of the Cartan-Weyl connection-type form of the Hilbert-Einstein action in the usual continuum GR.

In the Hamiltonian formalism, the tetrad bilinears are canonically conjugate to the orthogonal connection matrices. We evaluate the spectrum of the elementary areas, functions of the tetrad bilinears. The spectrum is discrete and proportional to the Faddeev analog $\gamma_{\rm F}$ of the Barbero-Immirzi parameter $\gamma$. The possibility of the tetrad and metric discontinuities in the Faddeev gravity allows to consider any surface as consisting of a set of virtually independent elementary areas and its spectrum being the sum of the elementary spectra. Requiring consistency of the black hole entropy calculations known in the literature we are able to estimate $\gamma_{\rm F}$.
\end{abstract}


MSC classes: 83C27; 53C05

keywords: Einstein theory of gravity; Faddeev gravity; piecewise flat spacetime; connection; area spectrum

\section{Introduction}

Faddeev formulation of gravity represents GR in terms of some extended set of tetrad variables which being excluded gives GR. Namely, a set of $d$ 4-vector (or four $d$-vector) fields $f^\lambda_A$ is considered \cite{Fad}. Here, Greek indices $\lambda, \mu$, \dots = 0, 1, 2, 3 label coordinates of the considered in the present paper pseudo-Riemannian spacetime and Latin capitals $A, B$, \dots = 0, \dots, $d - 1$ label coordinates of an external Minkowsky $d$-dimensional space. The metric tensor is
\begin{equation}                                                            
g_{\lambda\mu} = f^A_\lambda f_{\mu A}, ~~~ f_{\lambda A} = \eta_{AB} f^B_\lambda, ~~~ \eta = {\rm diag} (+1, \dots, +1, -1).
\end{equation}

\noindent Original choice is $d$ = 10, but further we discuss the case of an arbitrary $d$. %

The action is a functional of the field $f^A_\lambda$,
\begin{equation}\label{S}                                                   
S = \frac{1}{16 \pi G} \int \Pi^{AB} (f^\lambda_{A, \lambda} f^\mu_{B, \mu} - f^\lambda_{A, \mu} f^\mu_{B, \lambda}) \sqrt {- g} \d^4 x.
\end{equation}

\noindent Here, $f^\lambda_{A, \mu} \equiv \partial_\mu f^\lambda_A$, $\Pi_{AB} = \delta_{AB} - f^\lambda_A f_{\lambda B}$ is a projector which projects onto the so-called {\it vertical} subspace in the external space; the subspace orthogonal to the vertical one is called the {\it horizontal} subspace. Upon partial use of the equations of motion for $f^A_\lambda$ the expression (\ref{S}) becomes the Einstein action.

More generally, the Faddeev gravity has to do with gravity theories where the metric spacetime is not a fundamental physical concept but emerges from a non-spatio-temporal structure present in a more complete theory of interacting fundamental constituents (appearing, e. g., in the context of string theory) \cite{Chiu}.

A feature of the Faddeev action (\ref{S}) is that it remains finite for the discontinuous fields $f^A_\lambda$ (and therefore metrics $g_{\lambda \mu}$), since the action does not contain any of the squares of derivatives. In this regard, the theory is different from any of the usual field theories, including the usual GR. Despite the fact that this discontinuity does not survive on the equations of motion in classical framework (where, remind, this theory reduces to GR), the discontinuous metric is possible in the Faddeev gravity virtually in quantum framework.

This means that the metric on any 3d surface can be different if induced from the different two regions sharing this surface. Thus, the regions may not coincide geometrically on their common surface. To most clearly highlight this feature, we can use such a (minisuperspace) description that uses the division of space-time into regions, the interior metric degrees of freedom of which are frozen (elementary regions).

Such a minisuperspace formulation for the case of an ordinary GR is known and called Regge calculus \cite{Regge'}; see, e. g., review \cite{RegWil'}. In this approach, one considers the piecewise flat spacetimes, that is, the spacetimes composed of the flat 4D tetrahedra (4-simplices) $\sigma^4$ \cite{piecewiseflat=simplicial'}. Recent implementation of this approach, the Causal Dynamical Triangulations, has lead to important results in quantum gravity \cite{cdt}. As for the case of the Faddeev gravity, in particular, the 2d surface consists of virtually independent elementary areas. This may be important when analyzing the spectrum of the quantum surface area.

The quantization of area was discussed in the continuum theory as well, namely, using Ashtekar variables as early as in the work \cite{Ash}. Certain area operators in terms of the {\it discrete} Ashtekar type variables were considered in Ref. \cite{Loll}.

The spectrum of the surface area plays an important role in the black hole physics, in particular, in reproducing proportionality of the black hole entropy to its horizon area (Bekenstein-Hawking relation) by the statistical method.
If known, the spectrum of horizon area can be used to calculate the black hole entropy and find the condition that this entropy coincides with the Bekenstein-Hawking entropy. In the model of loop quantum gravity (LQG) based on Ashtekar variables \cite{ABCK} (see also Refs. \cite{LewDom,Mei}) this condition has allowed to find the Barbero-Immirzi parameter $\gamma$ \cite{Barb,Imm} to which the spectrum turns out to be proportional, for a simplified choice of the form of spectrum of the horizon area compared to the spectrum of the generic surface area. The Barbero-Immirzi parameter determines a term which can be added to the Cartan-Weyl form of the Hilbert-Einstein action and which vanishes on the equations of motion for the connection \cite{Holst,Fat} thus leading to the same Einstein action in terms of the metric only. Also there is a requirement of the so-called holographic bound principle for the entropy of any spherical nonrotating system including black hole \cite{Bek2,Tho,Sus}. To meet this requirement, in Refs \cite{Khr1,Glo,Khr2,Cor} the spectrum of the horizon area was chosen to coincide with the spectrum of the generic surface area, and the corresponding value of $\gamma$ found.

Returning to the opportunities provided by the minisuperspace Faddeev theory in finding the area spectrum, we have considered in our work \cite{II} this formulation in which our spacetime is piecewise flat. This is achieved by choosing the field $f^A_\lambda$ to be piecewise constant. This field is constant in the 4-simplices that make up the piecewise-flat manifold and, in turn, determines the lengths of their edges. Above possibility of the virtual discontinuity of $f^A_\lambda$ means that the edge lengths of the neighboring 4-simplices are considered as independent variables.

The elementary area tensor is some bilinear form of the fundamental discrete field variable $f^A_\lambda$. The most direct way to perform the canonical quantization on the basis of the Hamiltonian formalism is achieved by using the connection representation of the Faddeev gravity \cite{I} or, in the present context, its discrete analog \cite{III}. An analog of the Barbero-Immirzi parameter, $\gamma_{\rm F}$, is introduced there. Roughly speaking, elementary area is canonically conjugate to the connection variable which is an orthogonal matrix in the $d$-dimensional external spacetime. This fact leads to quantization of the elementary area in qualitative analogy with the way it happens with the quantization of angular momentum that is canonically conjugate to an orthogonal matrix of rotation in three-dimensional space. Namely, this quantization follows from the requirement that the wave function be single-valued w. r. t. the angle variables on which it depends. The elementary area turns out to be quantized as the momentum in some space with the dimension $d - 1$.

The concepts of the Lagrangian and canonical formalism suggest that one of the coordinates (time) is continuous. So the task is to perform a continuous time limit, assuming the distance between the vertices in the time direction tending to zero, pass to the canonical coordinates $p, q$ and analyze the emerging symplectic structure. Main formulas were derived already in our previous work \cite{Kha}, here we make the description more readable and compact and address a number of defining issues such as the canonical status of the variables used (Section \ref{background}), the question of the correspondence of the discrete first order Faddeev gravity to GR in the (nonperturbative) domain, where we have some reasonable area spectrum (one of the paragraphs after equation (\ref{sin-sin=0}) where some rotation $U_3$ not close to 1 is considered) and some others.
 
The paper is organized as follows. In Section \ref{kinetic}, the continuous time limit in the fully discrete action \cite{III} is performed and the kinetic part of the Lagrangian of the Faddeev discrete gravity (symbolically, $p\dot{q}$) is found. In Section \ref{background}, we specify the values of some non-dynamical variables parameterizing the connection (which cover the flat spacetime as well) other than the trivial ones and around which the area spectrum is well-defined and universal. In Section \ref{spectrum}, we find the spectrum of the elementary area, arising under the canonical quantization. The spectrum obtained is used to estimate the parameter $\gamma_{\rm F}$ using the equation, known in the literature, which expresses the condition that the statistical entropy of a black hole be equal to its Bekenstein-Hawking entropy. Thus, the Faddeev gravity can be consistent with black hole physics in this respect.

\section{ Time derivative term in the Lagrangian}\label{kinetic}

The Faddeev gravity action (\ref{S}) can be generalized by adding the $1/\gamma_{\rm F}$-term \cite{I},
\begin{equation}                                                            
S \Rightarrow \frac{1}{16 \pi G} \int \Pi^{AB} \left [ (f^\lambda_{A, \lambda} f^\mu_{B, \mu} - f^\lambda_{A, \mu} f^\mu_{B, \lambda}) \sqrt {- g} - \frac{1}{\gamma_{\rm F}} \epsilon^{\lambda \mu \nu \rho} f_{\lambda A, \mu} f_{\nu B, \rho} \right ] \d^4 x,
\end{equation}

\noindent and has the first order (connection) representation \cite{I}
\begin{eqnarray}\label{Faddeev-Cartan-Weyl}                                 
& & S (\omega, f) = \nonumber \\ & & \frac{1}{16 \pi G} \int \left \{ [ R_{\lambda \mu}^{AB} (\omega ) + \Lambda^\nu_{[\lambda \mu]} \omega_{\nu}^{AB} ] f^\lambda_A f^\mu_B \sqrt{- g} + \frac{ \epsilon^{\lambda \mu \nu \rho}}{2 \gamma_{\rm F}} f_{\nu A} f_{\rho B} R_{\lambda \mu}^{AB} (\omega ) \right \} \d^4 x, \nonumber \\
& & R_{\lambda \mu}^{ AB} (\omega ) = \partial_\lambda \omega_{\mu}^{ AB} - \partial_\mu \omega_{\lambda}^{ AB} + (\omega_\lambda \omega_\mu - \omega_\mu \omega_\lambda)^{AB}.
\end{eqnarray}

\noindent Here, $\Lambda^\nu_{[\lambda \mu]} \equiv - \Lambda^\nu_{[\mu \lambda]}$ are the coefficients at the constraints stating vanishing horizontal-horizontal block of the connection matrix $\omega_{\nu}^{AB}$, and $\gamma_{\rm F}$ is the direct analog  of the Barbero-Immirzi parameter.

In the discrete case, we consider the above representation for the manifold composed of hypercubes \cite{III},
\begin{eqnarray}\label{S-discr}                                             
& & \hspace{0mm} S^{\rm discr} = \frac{1}{16 \pi G} \sum_{\rm sites} \sum_{\lambda, \mu} \left\{ \frac{\sqrt{ (f^\lambda )^2 (f^\mu )^2 - (f^\lambda f^\mu )^2}}{ \sqrt{\det \| f^\nu f^\rho \|}} \right. \nonumber \\ & & \phantom{S^{\rm discr} =} \cdot \arcsin \left[ \frac{f^\lambda_A f^\mu_B - f^\mu_A f^\lambda_B}{2 \sqrt{ (f^\lambda )^2 (f^\mu )^2 - (f^\lambda f^\mu )^2}} R_{\lambda\mu}^{AB} (\Omega ) \right] \nonumber \\ & &  \hspace{0mm} \left. + \frac{1}{\gamma_{\rm F}} \sqrt{ \frac{(\epsilon^{\lambda \mu \nu \rho} f_{\nu A} f_{\rho B} )^2}{2}} \arcsin \left[ \frac{\epsilon^{\lambda \mu \nu \rho} f_{\nu A} f_{\rho B} }{ \sqrt{2 (\epsilon^{\lambda \mu \nu \rho} f_{\nu A} f_{\rho B}  )^2}} R_{\lambda\mu}^{AB} (\Omega ) \right] \right\} \nonumber \\ & & + \sum_{\rm sites} \sum_{\lambda, \mu, \nu} \Lambda^\lambda_{[\mu \nu]} \Omega^{AB}_\lambda (f^\mu_A f^\nu_B - f^\nu_A f^\mu_B), ~~~ R_{\lambda\mu} (\Omega ) = \dagOm_\lambda (T^{\rm T}_\lambda \dagOm_\mu) (T^{\rm T}_\mu \Omega_\lambda) \Omega_\mu .
\end{eqnarray}

\noindent Here, $T_\lambda$ is the translation operator along the edge $\lambda$ to the neighboring site; $f^\lambda_A$ (or $f_{\lambda A}$) and $\Omega^{AB}_\lambda$ are freely chosen vector and matrix field variables at the sites (vertices), that is, in hypercubes. The $d$-vector will be denoted in bold, like $\vf_\lambda$ for $f_{\lambda A}$. The matrix $\Omega$ as an element of SO(d-1,1) by default has one upper and one lower index, and
\begin{equation}                                                            
\dagOm = (\Omega \eta)^{\rm T} \eta.
\end{equation}

\noindent The last term in Eq. (\ref{S-discr}) represents certain additional condition on $\Omega_\lambda$, multiplied by the Lagrange multiplier $\Lambda^\lambda_{[\mu \nu]}$. This condition violates the local gauge SO(d-1,1) symmetry.

With metric discontinuities allowed, the piecewise constant {\it on hypercubes} metric can approximate (in a stepwise manner) any metric just as the piecewise constant {\it on simplices} metric can.

Let us pass to the continuous time. This procedure assumes that the 4-dimensional piecewise flat manifold is constructed of 3-dimensional piecewise flat manifolds (leaves) of the same structure labeled by a parameter $t$ (time), so that $t$ and $t \pm \d t$ label neighboring leaves and $\d t \to 0$. It is also assumed that
\begin{equation}                                                            
f^A_0 = O( \d t), ~~~ \Omega_0 = 1 + O( \d t)
\end{equation}

\noindent and the functions considered are differentiable so that
\begin{equation}                                                            
T_0 = 1 + \d t \frac{\d }{\d t} + O((\d t)^2).
\end{equation}

We have
\begin{equation}                                                            
R_{0 \lambda} - 1 = \dagOm_\lambda \dotOm_\lambda \d t + \dots , ~~~ - ( R_{\lambda 0} - 1 ) = \dagOm_\lambda \dotOm_\lambda \d t + \dots .
\end{equation}

\noindent Therefore the action
\begin{eqnarray}                                                           
& & \hspace{-10mm} S^{\rm discr} = \nonumber \\ & & \hspace{-10mm} \int \frac{\d t}{16 \pi G} \sum_{\rm sites} \sum_\lambda \left [ (f^0_A f^\lambda_B - f^\lambda_A f^0_B) \sqrt{- g} + \frac{1}{\gamma_{\rm F}} \epsilon^{0 \lambda \mu \nu} f_{\mu A} f_{\nu B} \right ] (\dagOm_\lambda \dotOm_\lambda)^{AB} + \dots .
\end{eqnarray}

\noindent Consider here contribution of a certain quadrangle, say, that one formed by $\vf_1$ and $\vf_2$ at a certain vertex. Let $\vnzero_1$, $\vnzero_2$ be a pair of mutually orthogonal unit vectors in the plane of $\vf_1$, $\vf_2$. Then
\begin{equation}                                                           
[\vf_1 \vf_2]_{AB} \equiv f_{1A} f_{2B} - f_{2A} f_{1B} = A [\vnzero_1 \vnzero_2]_{AB}
\end{equation}

\noindent where $A$ is the area of the quadrangle. Use the following identity,
\begin{equation}                                                           
( f^\lambda_A f^\mu_B - f^\mu_A f^\lambda_B )\sqrt{-g} = - \frac{1}{2} \epsilon^{\lambda \mu \nu \rho} \epsilon_{A B C D} f^C_\nu f^D_\rho,
\end{equation}

\noindent where $\epsilon_{A B C D}$ is the completely antisymmetric tensor in the horizontal subspace,
\begin{equation}                                                           
\epsilon_{A B C D} = \frac{\epsilon^{\lambda \mu \nu \rho} f_{\lambda A} f_{\mu B} f_{\nu C} f_{\rho D}}{\sqrt{-g}}.
\end{equation}

\noindent Then
\begin{equation}                                                           
\epsilon_{A B C D} n^{{\scriptscriptstyle (0)}C}_1 n^{{\scriptscriptstyle (0)}D}_2 = [\vnzero_0 \vnzero_3]_{AB}
\end{equation}

\noindent where $\vnzero_0, \vnzero_1, \vnzero_2, \vnzero_3$ form an orthonormal basis in the horizontal subspace,
\begin{equation}                                                           
\vnzero_\lambda \cdot \vnzero_\mu = 0 \mbox{~~~at~~~} \lambda \neq \mu, ~~~ (\vnzero_k)^2 = +1 \mbox{~~~at~~~} k = 1, 2, 3, ~~~ (\vnzero_0)^2 = -1.
\end{equation}

\noindent With these notations, contribution of the considered quadrangle to the kinetic term in the Lagrangian takes the form
\begin{equation}\label{L-kin}                                              
L = \frac{A}{16 \pi G} {\rm tr} \{ ( - \frac{1}{\gamma_{\rm F}} [\vnzero_1 \vnzero_2] + [\vnzero_0 \vnzero_3] ) \eta \dagOm_3 \dotOm_3 \}.
\end{equation}

Next we need an explicit ansatz for $\Omega$. The constraint $\partial S^{\rm discr} / \partial \Lambda = 0$ on $\Omega$ as it is given in $S^{\rm discr}$ is a matrix transcendental equation (on the generator of $\Omega$) and cannot be explicitly solved. However, this constraint is allowed to be modified somehow \cite{III} if there is no change in the leading order in the generator of $\Omega$. Besides that, the contribution of the vertical-vertical block of $\omega^{AB}_\nu$ to the action in the continuum case identically vanishes. However, in the discrete formulation, the vertical-vertical part of $\Omega$ contributes nonzero value to the action due to the more complicated nonlinear structure of the latter. Therefore, there is some ambiguity when passing to the discrete theory from the continuum formulation depending on whether we impose some constraints on the vertical-vertical components of the connection or not. Now we choose to impose such constraints and to put vertical-vertical part of $\omega^{AB}_\nu$ to be zero. That is, in overall,
\begin{equation}\label{Omega-ff-Omega-PiPi-zero}                           
\omega_{\nu}^{AB} f^\lambda_A f^\mu_B = 0, ~~~ \omega_{\nu}^{AB} \Pi_{AC} \Pi_{BD} = 0.
\end{equation}

\noindent This has the general solution
\begin{equation}\label{omega=fb}                                           
\omega_{\lambda AB} = f^\mu_A b_{\mu \lambda B} - f^\mu_B b_{\mu \lambda A} \equiv [ \vf^\mu \vb_{\mu \lambda} ]_{AB}, ~~~ \vb_{\mu \lambda } \cdot \vf_\nu = 0
\end{equation}

\noindent parameterized by a new independent vertical vector variable $\vb_{\mu \lambda}$. Upon substituting this to the action we can find $b_{\mu \lambda A}$ from the equations of motion for it,
\begin{equation}                                                           
b_{\mu \lambda A} = f_{\mu B, \lambda} \Pi^B_A.
\end{equation}

\noindent We can view $\vf^\mu$ in (\ref{omega=fb}) being expanded over a set of some orthonormal frame vectors $\vnzero_\mu$ and get analogous expression for $\omega_\lambda$ with some vertical vector variable $\tilde{\vb}_{\mu \lambda}$ (a combination of $\vb_{\mu \lambda}$s),
\begin{equation}                                                           
\omega_{\lambda} = \sum_\mu [ \vnzero_\mu \tilde{\vb}_{\mu \lambda} ] \eta .
\end{equation}

In the discrete case, we can transfer this to (the antisymmetric part of) $\Omega_3$ of interest,
\begin{equation}\label{Omega-dagOmega=nb}                                  
\frac{1}{2} ( \Omega_3 - \dagOm_3 ) = \sum_\lambda [ \vnzero_\lambda \vb_\lambda ] \eta + \dots,
\end{equation}

\noindent where $\vb_\lambda \equiv \vb_{\lambda 3}$, tildes being omitted. Here, the dots mean the terms of the larger orders over $\vb_\lambda$ (these terms, among other things, do not contribute to the action in the continuum limit). Consider restoring full nonlinear $\Omega_3$ as
\begin{equation}\label{Omega=uuuu}                                         
\Omega_3 = u_2 u_0 u_3 u_1, ~~~ u_\lambda = \exp ( [ \vn_\lambda \vb_\lambda ] \eta ).
\end{equation}

\noindent Here, $\vb$ is orthogonal to some orthonormal set $\vn$ and generally $\vn_\lambda$ can differ from $\vnzero_\lambda$ by an amount tending to zero if $\vb_\mu$s tend to zero. Dragging $u_0$ through $u_2$ and $u_3$ through $u_1$ rewrite $\Omega_3$ as
\begin{eqnarray}\label{Omega=uuuu-new}                                     
& & \Omega_3 = u_{0(2)} u_2 u_1 u_{3(1^+)}, \\ & & u_{0(2)} = u_2 u_0 \dagu_2 = \exp ( [ \vn_0 \vb_{0(2)} ] \eta ), ~~~ \vb_{0(2)} = u_2 \vb_0, \nonumber\\ & & u_{3(1^+)} = \dagu_1 u_3 u_1 = \exp ( [ \vn_3 \vb_{3(1^+)} ] \eta ), ~~~ \vb_{3(1^+)} = \dagu_1 \vb_3. \nonumber
\end{eqnarray}

\noindent Let us parameterise $\vnzero$ in terms of $\vn$ as $\vnzero_\lambda = \dagu_1 \dagu_3 \vn_\lambda$. This implicitly defines $\vn$ in terms of $\vnzero, \vb_1, \vb_3$. Then $u_1 [ \vnzero_1 \vnzero_2] \eta \dagu_1 = [ \vn_1 \vn_2] \eta$, $u_{3(1^+)} [\vnzero_0 \vnzero_3] \eta \dagu_{3(1^+)} = [\vn_0 \vn_3] \eta$. Substitute Eq (\ref{Omega=uuuu}) to the $1/\gamma_{\rm F}$ term in (\ref{L-kin}), and Eq (\ref{Omega=uuuu-new}) to the main $(\gamma_{\rm F})^0$ term. Simple calculation gives
\begin{eqnarray}\label{L-kin-ansatz}                                       
& & \frac{16 \pi G L}{A} = - \frac{1}{\gamma_{\rm F}} {\rm tr} [ \vn_1 \vn_2] \eta (\dagu_2 \dotu_2 + \dotu_1 \dagu_1 ) + {\rm tr} [\vn_0 \vn_3] \eta ( \dagu_{0(2)} \dotu_{0(2)} + \dotu_{3(1^+)} \dagu_{3(1^+)} ), \\ & & u_\lambda = \exp ( [ \vn_\lambda \vb_\lambda ] \eta ), \nonumber \\ & & u_{0(2)} = \exp ( [ \vn_0 \vb_{0(2)} ] \eta ), ~~ \vb_{0(2)} = u_2 \vb_0, ~~ u_{3(1^+)} = \exp ( [ \vn_3 \vb_{3(1^+)} ] \eta ), ~~ \vb_{3(1^+)} = \dagu_1 \vb_3. \nonumber
\end{eqnarray}

\noindent It is not quite trivial that we could partially disentangle variables on the nonlinear level correcting the constraint term in an admissible way.

The exact nonlinear form of the adopted ansatz for connection is quite simple, e. g. for the Euclidean rotation ($\vn^2 = +1$)
\begin{eqnarray}                                                           
& & \exp ([ \vn \vb ] \eta ) \eta = \eta + ( \vn \otimes \vn + \vl \otimes \vl ) ( \cos b - 1) + [\vn \vl ] \sin b, \\ & & b = \sqrt{\vb^2}, ~~~ \vl = \vb / b, ~~~ \vl_\0 = \vb_{0(2)} / b_0, ~~~ \vl_\3 = \vb_{3(1^+)} / b_3. \nonumber
\end{eqnarray}

\noindent This allows to rewrite Eq. (\ref{L-kin-ansatz}) as
\begin{eqnarray}\label{L-kin-final}                                        
\frac{8 \pi G L}{A} = - \frac{1}{\gamma_{\rm F}} \left [ (\vn_2 \cdot \dotvn_1) (\cos b_2 - \cos b_1 ) + (\vl_1 \cdot \dotvn_2 ) \sin b_1 - (\vl_2 \cdot \dotvn_1 ) \sin b_2 \right ] \nonumber \\ - ( \vn_0 \cdot \dotvn_3 ) ( \cosh b_0 - \cos b_3 ) - ( \vl_\3 \cdot \dotvn_0 )\sin b_3 - (\vl_\0 \cdot \dotvn_3 ) \sinh b_0.
\end{eqnarray}

\noindent Here, if summarized, $\vn, \vl$ are new $d$-vector variables and $b$ are scalar variables, independent, except for the fact that $\vl$ are unit vectors orthogonal to $\vn_0, \vn_1, \vn_2, \vn_3$ forming an orthonormal set.

\section{ Proper background configuration}\label{background}

To get an idea of the canonical status of some variables, compare the above with the continuum action $S (\omega, f)$ where $\omega$ in terms of $b$ (\ref{omega=fb}) is taken,
\begin{eqnarray}                                                           
& & S_b (b, f) = \frac{1}{16 \pi G} \int \left \{ \left [ 2 f^\mu_A \left ( b^{\lambda A}_{~ \mu, \lambda} - b^{\lambda A}_{~ \lambda, \mu} \right ) + b^{\lambda A}_{~\mu} b^\mu_{~\lambda A} - b^{\lambda A}_{~\lambda} b^\mu_{~\mu A} \right ] \sqrt{- g} \right. \nonumber \\ & & \left. + \frac{ \epsilon^{\lambda \mu \nu \rho}}{ \gamma_{\rm F}} \left ( 2 f_{\lambda A} b^A_{\nu \rho , \mu} + b_{\lambda \mu A} b^A_{\nu \rho} \right ) \right \} \d^4 x, ~~~ b^A_{\lambda \mu} f^\nu_A = 0.
\end{eqnarray}

\noindent Separating the space and time indices we get
\begin{eqnarray}                                                           
& & S_b (b, f) = \frac{1}{8 \pi G} \int \left [ \dot{f}^k_A p^A_k + \dot{f}^0_A p^A \right. \nonumber \\ & & \left. + b^k_{~ 0A} \left ( - p^A_k - f^{0A}_{,k} + \frac{g_{ki} \epsilon^{ilm}}{\gamma_{\rm F} \sqrt{-g}} f^A_{l,m} \right ) + b^0_{~ 0A} \left ( - p^A + f^{kA}_{,k} + \frac{g_{0i} \epsilon^{ilm}}{\gamma_{\rm F} \sqrt{-g}} f^A_{l,m} \right ) \right. \nonumber \\ & & \left. + f^k_A \left ( b^{lA}_{~ k,l} - b^{lA}_{~ l,k} \right ) + \frac{1}{2} \left ( b^{kA}_{~ l} b^l_{~ kA} - b^{kA}_{~ k} b^l_{~ lA} \right ) \right. \nonumber \\ & & \left. + \frac{\epsilon^{ikl}}{\gamma_{\rm F} \sqrt{-g}} \left ( - f_{kA} b^A_{0l,i} + f_{0A} b^A_{ik,l} + b^A_{0l} b_{ikA} \right ) \right ] \sqrt{- g} \d^4 x.
\end{eqnarray}

\noindent Here, some combinations of $b^{\lambda A}_{~ \mu}$ play the role of canonical momenta,
\begin{eqnarray}                                                           
p^A_k = - b^{0A}_{~ k} + \frac{g_{ki} \epsilon^{ilm}}{\gamma_{\rm F} \sqrt{-g}} b^A_{lm}, ~~~ p^A = b^{kA}_{~ k} + \frac{g_{0i} \epsilon^{ilm}}{\gamma_{\rm F} \sqrt{-g}} b^A_{lm}.
\end{eqnarray}

\noindent Varying the action over $b^k_{~ 0A}$, $b^0_{~ 0A}$ results in some constraints on $p^A_k$, $p^A$ (which here simply express these in terms of the canonical coordinates $f^\lambda_A$). The other non-dynamical $b^{\lambda A}_{~ \mu}$ can be found in terms of $p$, $q$. In the discrete version, analogs of $b^k_{~ 0A}$, $b^0_{~ 0A}$ enter nonlinearly, and analogs of $\delta S_b / \delta b^k_{~ 0A} = 0$, $\delta S_b / \delta b^0_{~ 0A} = 0$ do not result in the constraints but rather allow to define the analogs of $b^k_{~ 0A}$, $b^0_{~ 0A}$ themselves in terms of $p$, $q$. Discrete analogs of the canonical momenta remain unconstrained.

In the discrete kinetic term (\ref{L-kin-final}), the variables $\vb$ (or $b$, $\vl$) are not subject to the constraints (other than the orthogonality conditions listed after Eq. (\ref{L-kin-final})). Besides that, the variables $b$ are non-dynamical for their velocities can not be defined from the equations of motion.

Now, performing a series of elementary transformations of the variables $\vb$, $\vn$, we can block-diagonalize this symplectic form on the phase space. First consider the purely $1/\gamma_{\rm F}$-part. Let us represent $\vl_1$, $\vl_2$ in terms of orthogonal vectors,
\begin{eqnarray}                                                           
& & \vl_1 = \lambda_1 \ve_1 + \lambda_2 \ve_2, ~~~ \vl_2 = \lambda_1 \ve_1 - \lambda_2 \ve_2, \\ & & \ve_1^2 = \ve_2^2 = 1, ~~~ \ve_1 \cdot \ve_2 = 0, ~~~ \lambda_{1,2} = \sqrt{\frac{1 \pm \vl_1 \cdot \vl_2 }{2}}. \nonumber
\end{eqnarray}

\noindent This gives a combination of the terms $\vn_2 \cdot \dotvn_1$, $\ve_1 \cdot \dotvn_2$, $\ve_2 \cdot \dotvn_2$, $\ve_1 \cdot \dotvn_1$, $\ve_2 \cdot \dotvn_1$. Let us pass to new field variables by rotating in the 2-planes of $\vn_1, \vn_2$ and $\ve_1, \ve_2$,
\begin{equation}                                                           
\left. \begin{array}{c}
\vn_1 = \prvn_1 \cos \alpha - \prvn_2 \sin \alpha \\
\vn_2 = \prvn_1 \sin \alpha + \prvn_2 \cos \alpha
\end{array} \right\} , ~~~
\left. \begin{array}{c}
\ve_1 = \phantom{-} \prve_1 \cos \beta + \prve_2 \sin \beta \\
\ve_2 = - \prve_1 \sin \beta + \prve_2 \cos \beta
\end{array} \right\} .
\end{equation}

\noindent Let us choose $\alpha, \beta$ to cancel the terms with $\prve_1 \cdot \prdotvn_1$ and $\prve_2 \cdot \prdotvn_2$. Next rotate in the 2-planes of $\prvn_1, \prve_1$ and $\prvn_2, \prve_2$,
\begin{equation}                                                           
\left. \begin{array}{c}
\prvn_i = \pprvn_i \cos \alpha_i - \pprve_i \sin \alpha_i \\
\prve_i = \pprvn_i \sin \alpha_i + \pprve_i \cos \alpha_i
\end{array} \right\} i = 1, 2,
\end{equation}

\noindent choosing $\alpha_1, \alpha_2$ to cancel the terms with $\pprve_2 \cdot \pprdotvn_1$ and $\pprve_1 \cdot \pprdotvn_2$. The result reads
\begin{eqnarray}\label{L-gamma}                                            
& & - \gamma_{\rm F} \frac{8 \pi G}{A} L_\gamma = \frac{(\cos b_2 - \cos b_1)(\sin^2 b_1 - \sin^2 b_2 )^2}{(\sin^2 b_1 - \sin^2 b_2 )^2 + 4 (\vl_1 \cdot \vl_2 )^2 \sin^2 b_1 \sin^2 b_2 } ~ \frac{\d}{\d t} ~ \frac{ (\vl_1 \cdot \vl_2 ) \sin b_1 \sin b_2 }{ \sin^2 b_1 - \sin^2 b_2 } \nonumber \\
& & \phantom{- \gamma_{\rm F} \frac{8 \pi G}{A} L =} + \frac{1}{2} \left ( \sqrt{2 - 2 \cos b_1 \cos b_2 + 2 \llangle \sin b_1 \sin b_2} \right. \nonumber \\ & & \phantom{- \gamma_{\rm F} \frac{8 \pi G}{A} L =} \left. + \sqrt{2 - 2 \cos b_1 \cos b_2 - 2 \llangle \sin b_1 \sin b_2} \right ) \pprvn_2 \cdot \pprdotvn_1 \nonumber \\
& & \phantom{- \gamma_{\rm F} \frac{8 \pi G}{A} L =} + \frac{1}{2} \left ( \sqrt{2 - 2 \cos b_1 \cos b_2 + 2 \llangle \sin b_1 \sin b_2} \right. \nonumber \\ & & \phantom{- \gamma_{\rm F} \frac{8 \pi G}{A} L =} \left. - \sqrt{2 - 2 \cos b_1 \cos b_2 - 2 \llangle \sin b_1 \sin b_2} \right ) \pprve_2 \cdot \pprdotve_1
\end{eqnarray}

\noindent The area operator can be formed on the basis of the (angle type) canonical variables related to the pair either $\pprvn_2, \pprvn_1$ or $\pprve_2, \pprve_1$ and should have a discrete spectrum. The consistency of the spectrum requires that either coefficients at $\pprvn_2 \cdot \pprdotvn_1$ and $\pprve_2 \cdot \pprdotve_1$ be equal in absolute value or one of these be zero. This imposes certain conditions on $\vb$. The least restrictive of these is $b_1 = \pi$ or $b_2 = \pi$ which singles out some configuration subspace of codimension 1. This is a particular solution to
\begin{equation}                                                           
\llangle \sin b_1 \sin b_2 = 0
\end{equation}

\noindent which is necessary for the coefficient at $\pprve_2 \cdot \pprdotve_1$ being zero. The other solutions correspond to subspaces of larger codimension (smaller dimension) and should be discarded according to the probabilistic consideration. For example, the solution $b_1 = 0$ (or $b_2 = 0$) means $\vb_1 = 0$ (or $\vb_2 = 0$) and thus defines a subspace of the codimension $d - 4$ (the dimensionality of the vertical subspace where $\vb_1$ or $\vb_2$ can vary). The condition $\vl_1 \cdot \vl_2 = 0$, $b_1 = b_2$ which provides equality of the coefficients at $\pprvn_2 \cdot \pprdotvn_1$ and $\pprve_2 \cdot \pprdotve_1$ singles out a subspace of codimension 2, and so on. So let us take $b_1 = \pi$ for definiteness.

Next consider the purely $(\gamma_{\rm F})^0$ part. Let us represent $\vl_\3$, $\vl_\0$ in terms of orthogonal vectors,
\begin{eqnarray}                                                           
& & \vl_\3 = \lambda_3 \ve_3 + \lambda_0 \ve_0, ~~~ \vl_\0 = \lambda_3 \ve_3 - \lambda_0 \ve_0, \\ & & \ve_3^2 = \ve_0^2 = 1, ~~~ \ve_3 \cdot \ve_0 = 0, ~~~ \lambda_{3,0} = \sqrt{\frac{1 \pm \vl_\3 \cdot \vl_\0 }{2}}. \nonumber
\end{eqnarray}

\noindent Thus we arrive at a combination of the terms $\vn_0 \cdot \dotvn_3$, $\ve_3 \cdot \dotvn_0$, $\ve_0 \cdot \dotvn_0$, $\ve_3 \cdot \dotvn_3$, $\ve_0 \cdot \dotvn_3$. Let us rotate in the 2-planes of $\vn_3, \vn_0$ and $\ve_3, \ve_0$,
\begin{equation}                                                           
\left. \begin{array}{c}
\vn_3 = \prvn_3 \cosh \xi + \prvn_0 \sinh \xi \\
\vn_0 = \prvn_1 \sinh \xi + \prvn_0 \cosh \xi
\end{array} \right\} , ~~~
\left. \begin{array}{c}
\ve_3 = \phantom{-} \prve_3 \cos \zeta + \prve_0 \sin \zeta \\
\ve_0 = - \prve_3 \sin \zeta + \prve_0 \cos \zeta
\end{array} \right\} .
\end{equation}

\noindent Let us choose $\zeta, \xi$ to cancel the terms with $\prve_3 \cdot \prdotvn_3$ and $\prve_0 \cdot \prdotvn_0$. Next rotate in the 2-planes of $\prvn_3, \prve_3$ and $\prvn_0, \prve_0$,
\begin{equation}                                                           
\left. \begin{array}{c}
\prvn_3 = \pprvn_3 \cos \alpha_3 - \pprve_3 \sin \alpha_3 \\
\prve_3 = \pprvn_3 \sin \alpha_3 + \pprve_3 \cos \alpha_3
\end{array} \right\} , ~~~
\left. \begin{array}{c}
\prvn_0 = \pprvn_0 \cosh \xi_0 + \pprve_0 \sinh \xi_0 \\
\prve_0 = \pprvn_0 \sinh \xi_0 + \pprve_0 \cosh \xi_0
\end{array} \right\} ,
\end{equation}

\noindent choosing $\alpha_3, \xi_0$ to cancel the terms with $\pprve_0 \cdot \pprdotvn_3$ and $\pprve_3 \cdot \pprdotvn_0$. The result reads
\begin{eqnarray}\label{L-0}                                                
& & - \frac{8 \pi G}{A} L_0 = \frac{(\cosh b_0 - \cos b_3)(\sinh^2 b_0 + \sin^2 b_3 )^2}{(\sinh^2 b_0 + \sin^2 b_3 )^2 - 4 (\vl_\3 \cdot \vl_\0 )^2 \sin^2 b_3 \sinh^2 b_0 } ~ \frac{\d}{\d t} ~ \frac{ (\vl_\3 \cdot \vl_\0 ) \sin b_3 \sinh b_0 }{ \sinh^2 b_0 + \sin^2 b_3 } \nonumber \\
& & \phantom{- \frac{8 \pi G}{A} L =} + \frac{1}{2} \left ( \sqrt{2 - 2 \cos b_3 \cosh b_0 + 2 i \nllangle \sin b_3 \sinh b_0} \right. \nonumber \\ & & \phantom{- \frac{8 \pi G}{A} L =} \left. + \sqrt{2 - 2 \cos b_3 \cosh b_0 - 2 i \nllangle \sin b_3 \sinh b_0} \right ) \pprvn_0 \cdot \pprdotvn_3 \nonumber \\
& & \phantom{- \frac{8 \pi G}{A} L =} + \frac{1}{2 i} \left ( \sqrt{2 - 2 \cos b_3 \cosh b_0 + 2 i \nllangle \sin b_3 \sinh b_0} \right. \nonumber \\ & & \phantom{- \frac{8 \pi G}{A} L =} \left. - \sqrt{2 - 2 \cos b_3 \cosh b_0 - 2 i \nllangle \sin b_3 \sinh b_0} \right ) \pprve_0 \cdot \pprdotve_3 .
\end{eqnarray}

\noindent The area operator formed on the basis of the (angle type) canonical variables related to the pair $\pprve_0, \pprve_3$ should have a discrete spectrum. (The pair $\pprvn_0, \pprvn_3$ does not lead to the discrete spectrum since $\pprvn_0$ varies in the noncompact region.) Consistency of the spectrum with the above description on the basis of the pairs appearing in $L_\gamma$ requires
\begin{equation}\label{sin-sin=0}                                          
\nllangle \sin b_3 \sinh b_0 = 0.
\end{equation}

\noindent Again, the solution $b_3 = \pi$ provides us with the configuration subspace of the largest dimensionality.

The values $b_1 = \pi$, $b_3 = \pi$ found mean only $\pprvn_2 \cdot \pprdotvn_1$, $\pprvn_0 \cdot \pprdotvn_3$ terms in $L$ and the absence of $\pprve_2 \cdot \pprdotve_1$, $\pprve_0 \cdot \pprdotve_3$ terms.

Normally, if the piecewise flat spacetime is close to the flat spacetime or approximates a smooth spacetime arbitrarily accurately, the curvature matrix is arbitrarily close to unity. Usually, $\Omega$s are taken close to unity for that ($b_\lambda \ll 1$). Then the area operator formed on the basis of any suitable subset of the (real angle type) canonical variables in $L$ will have a discrete spectrum proportional to $\varepsilon^{-1}$ where $\varepsilon$ is a typical scale of $b_\lambda$. This dependence is singular at $\varepsilon \to 0$ and makes the spectrum not universal. However, the curvature matrix $\dots (T^{\rm T} \dagOm) \dots \Omega \dots$ can be close to unity if $\Omega$ varies slowly from site to site even if it comprises reflections (some $b_\lambda$ are $\pi$).

We just have seen that the consistent universal spectrum can arise upon analytic continuation of the values of some $b_\lambda$s to $\pi$. In our example, $\Omega_3$ at a vertex $(x_0^1, x_0^2, x_0^3, x_0^0)$ is close to $U_3 = \exp (\pi [\vn_3 \vl_3]\eta ) \exp (\pi [\vn_1 \vl_1]\eta )$. Suppose $\Omega_3$ at the next site along the $x_3$ axis is close to $U_3^\dagger |_{x^3 = x_0^3} = \exp (\pi [\vn_1 \vl_1]\eta ) \exp (\pi [\vn_3 \vl_3]\eta ) |_{x^3 = x_0^3}$, at $x^3 = x_0^3 + 2$ again to $U_3 |_{x^3 = x_0^3 + 2}$ and so on. This could be interpreted as frame rotations (reflections) by $U$ in the cubes of the layers $x^3 = const$ through one (say, with odd $x^3$). We could check whether the theory around the proposed background connection could be classically equivalent to GR. For that, we could redefine the tetrad by $U$ in the layers $x^3 = const$ through one, $U \vf^\lambda \equiv \vg^\lambda$ in the odd layers and $\vf^\lambda \equiv \vg^\lambda$ in the even layers. Assuming that the redefined tetrad $\vg^\lambda$ varies slowly from site to site, we could check whether the full discrete connection action is equivalent to the Faddeev action on $\vg^\lambda$ upon excluding the connection in the continuum limit, that is, whether it is equivalent to GR action with $g_{\lambda \mu} = \vf_\lambda \cdot \vf_\mu = \vg_\lambda \cdot \vg_\mu$. The situation is more complicated than simply the gauge transformation by $U$ because the Faddeev gravity in the connection representation is not locally gauge invariant (since the constraints $\delta S / \delta \Lambda^\nu_{[\lambda \mu]} = 0$ or (\ref{Omega-ff-Omega-PiPi-zero}) are not), but the answer is yes. Checking this is the subject of a separate publication.

Analogously to the bivector $[\vf_1 \vf_2]$, we could consider the bivectors $[\vf_2 \vf_3]$ and $[\vf_3 \vf_1]$. The variables related to these are not independent. As a result, we can define the spectrum for only one area per cuboid at the same time. A priori the possibility that the spectrum of any of these areas will be consistent is provided by $\Omega_1$, $\Omega_2$ containing the reflections (as $\exp (\pi [\vn \vl]\eta )$ in $\Omega_3$) related to $f^A_\lambda$. Then the curvature $R_{\lambda\mu} (\Omega )$ generally will be different from unity already for constant $f^A_\lambda$, and establishing conditions for reproducing classically the GR action around the proposed background connection requires a deeper analysis.

Returning to the spectrum for the bivector  $[\vf_1 \vf_2]$, if we consider not only strict equality of some $b_\lambda$ to $\pi$, but the whole neighborhood of $\pi$, $(b_1, b_2, b_3, b_0) = (\pi + \varepsilon_1, \varepsilon_2, \pi + \varepsilon_3, \varepsilon_0)$ in our example, then the terms $\pprve_2 \cdot \pprdotve_1$, $\pprve_0 \cdot \pprdotve_3$ are present, but their coefficients in Eqs (\ref{L-gamma}) and (\ref{L-0}) are $O ( \varepsilon^2 )$. Also the non-constant parts of the coefficients at $\pprvn_2 \cdot \pprdotvn_1$, $\pprvn_0 \cdot \pprdotvn_3$ are $O( \varepsilon^2 )$. Besides that, whereas $\pprvn_2 \cdot \pprvn_1 = 0 = \pprvn_0 \cdot \pprvn_3$, the scalar products of $\pprvn_2, \pprvn_1$ with $\pprvn_0, \pprvn_3$ are $O(\varepsilon^2)$. Therefore, some orthonormal set $\ppprvn_\lambda = \pprvn_\lambda + O( \varepsilon^2 )$ exists. (We have up to $O ( \varepsilon^2 )$: $\ppprvn_1 = \vn_1 + \varepsilon_1 \vl_1 / 2$, $\ppprvn_2 = \vn_2 - \varepsilon_2 \vl_2 / 2$, $\ppprvn_3 = \vn_3 + \varepsilon_3 \vl_\3 / 2$, $\ppprvn_0 = \vn_0 + \varepsilon_0 \vl_\0 / 2$ and, in turn, $\vn_1 = - \vnzero_1 - \varepsilon_1 \vl_1$, $\vn_2 = \vnzero_2$, $\vn_3 = - \vnzero_3 - \varepsilon_3 \vl_\3$, $\vl_\3 = \vl_3 - 2 (\vl_1 \cdot \vl_3 ) \vl_1 + O( \varepsilon )$, $\vn_0 = \vnzero_0$.)

The bilinears in $\varepsilon_\lambda$ are of the order of some contribution to the curvature matrix and are therefore $O ( a^2 R )$ (or, in any case, do not exceed it) where $a$ is a typical edge length and $R$ is a typical curvature of the smooth geometry which is approximated by the considered piecewise flat one. The $O ( \varepsilon^2 )$ terms in $L$ are $O (A a^2 R ) = O( a^4 R )$. Since the number of the cubes with the edge length $a$ in the given volume $V$ in the 3-dimensional section is $V / a^3$, the contribution of these terms to $L$ in this volume is $O (a )$ and thus tends to zero in the continuum limit $a \to 0$.

Thus, if we allow a modification of the discrete action (adding terms that vanish in the continuum limit), $L$ can be (upon passing to $\ppprvn_\lambda$ and omitting the primes)
\begin{equation}                                                           
L = - \frac{A}{4 \pi G} \left ( \frac{1}{\gamma_{\rm F}} \vn_2 \cdot \dotvn_1 + \vn_0 \cdot \dotvn_3 \right ).
\end{equation}

\noindent The area in the plane $x^1, x^2$ is considered and $\vn_\lambda$ form an orthonormal system.

\section{ Elementary area spectrum}\label{spectrum}

Defining area spectrum is then straightforward. Let us parameterize the vectors $\vn_\lambda$ successively with the help of the (pseudo-)spherical coordinates (angle part), an analog of the Euler angles. Let us start with $\vn_0$,
\begin{equation}                                                           
\vn_0 = \left ( \begin{array}{l}
\sinh \chi_{d-1} \sin \chi_{d-2} \dots \sin \chi_1 \\
\sinh \chi_{d-1} \sin \chi_{d-2} \dots \cos \chi_1 \\
\vdots \\
\cosh \chi_{d-1}
\end{array} \right ).
\end{equation}

\noindent The orthonormal basis in the $(d-1)$-dimensional orthogonal subspace is naturally chosen as
\begin{eqnarray}                                                           
& & \vn^{(1)}_1 = \left ( \begin{array}{l}
\phantom{-} \cos \chi_1 \\
- \sin \chi_1 \\
\phantom{-} 0 \\
\phantom{-} \vdots \\
\phantom{-} 0
\end{array} \right ) , ~~~
\vn^{(2)}_1 = \left ( \begin{array}{l}
\phantom{-} \cos \chi_2 \sin \chi_1 \\
\phantom{-} \cos \chi_2 \cos \chi_1 \\
- \sin \chi_2 \\
\phantom{-} \vdots \\
\phantom{-} 0
\end{array} \right ) , \dots , \nonumber \\
& & \vn^{(d-1)}_1 = \left ( \begin{array}{l}
\phantom{} \cosh \chi_{d-1} \sin \chi_{d-2} \dots \sin \chi_1 \\
\phantom{} \cosh \chi_{d-1} \sin \chi_{d-2} \dots \cos \chi_1 \\
\phantom{} \vdots \\
\phantom{} \vdots \\
\sinh \chi_{d-1}
\end{array} \right ).
\end{eqnarray}

\noindent Then
\begin{equation}                                                           
\vn_1 = D_1 \vn^{(1)}_1 + \dots +D_{d-1} \vn^{(d-1)}_1, ~~~ D^2_1 + \dots +D^2_{d-1} = 1
\end{equation}

\noindent and the $(d - 1)$-dimensional vector $\tvn_1 = (D_1, D_2, \dots, D_{d - 1})^{\rm T}$ is $\vn_1$ in the basis $\vn^{(1)}_1, \vn^{(2)}_1$, \dots, $\vn^{(d-1)}_1$. This procedure of transition from $\vn_0$ to $\tvn_1$ is repeated and applied to $\tvn_1$ and then to $\tvn_2$ to give, in overall, the transition
\begin{eqnarray}                                                           
& & \vn_0 \Longrightarrow \tvn_1 \Longrightarrow \tvn_2 \Longrightarrow \tvn_3, \nonumber \\
& & \vn_2 = B_1 \vn^{(1)}_2 + \dots +B_{d-2} \vn^{(d-2)}_2, ~~~ B^2_1 + \dots + B^2_{d-2} = 1, \nonumber \\
& & \vn_3 = C_1 \vn^{(1)}_3 + \dots + C_{d-3} \vn^{(d-3)}_3, ~~~ C^2_1 + \dots + C^2_{d-3} = 1.
\end{eqnarray}

\noindent The difference from $\vn_0$ is that $\tvn_1$ and $\tvn_2$ are parameterized by spherical coordinates $\theta_1, \theta_2, \dots, \theta_{d - 2}$ and $\varphi_1, \varphi_2, \dots, \varphi_{d - 3}$, respectively, rather than by {\it pseudo}spherical ones as $\chi$. For example,
\begin{equation}                                                           
\tvn_1 \equiv \left ( \begin{array}{l}
D_1 \\
D_2 \\
\vdots \\
D_{d-1}
\end{array} \right ) = \left ( \begin{array}{l}
\sin \theta_{d-2} \sin \theta_{d-3} \dots \sin \theta_1 \\
\sin \theta_{d-2} \sin \theta_{d-3} \dots \cos \theta_1 \\
\vdots \\
\cos \theta_{d-2}
\end{array} \right ),
\end{equation}

\noindent and the orthonormal basis in the $(d-2)$-dimensional orthogonal subspace is naturally chosen as
\begin{eqnarray}                                                           
& & \vn^{(1)}_2 = \left ( \begin{array}{l}
\phantom{-} \cos \theta_1 \\
- \sin \theta_1 \\
\phantom{-} 0 \\
\phantom{-} \vdots \\
\phantom{-} 0
\end{array} \right ) , ~~~
\vn^{(2)}_2 = \left ( \begin{array}{l}
\phantom{-} \cos \theta_2 \sin \theta_1 \\
\phantom{-} \cos \theta_2 \cos \theta_1 \\
- \sin \theta_2 \\
\phantom{-} \vdots \\
\phantom{-} 0
\end{array} \right ) , \dots , \nonumber \\
& & \vn^{(d-2)}_2 = \left ( \begin{array}{l}
\phantom{-} \cos \theta_{d-2} \sin \theta_{d-3} \dots \sin \theta_1 \\
\phantom{-} \cos \theta_{d-2} \sin \theta_{d-3} \dots \cos \theta_1 \\
\phantom{-} \vdots \\
\phantom{-} \vdots \\
- \sin \theta_{d-2}
\end{array} \right ),
\end{eqnarray}

\noindent so that $\vn_2 = B_1 \vn^{(1)}_2 + \dots +B_{d-2} \vn^{(d-2)}_2$.

The row of column vectors $( \vn^{(1)}_k \vn^{(2)}_k \dots \vn^{(d-k)}_k )$ is a $(d - k + 1) \times (d - k)$ matrix $M_k, k = 1, 2, 3$. By definition, these possess the properties $M_1^{\rm T} \eta \vn_0 = 0$, $M_k^{\rm T} \tvn_{k - 1} = 0$ at $k = 2, 3$, $M^{\rm T}_1 \eta M_1 = 1$ (the $(d - 1) \times (d - 1)$ matrix) or $M^{\rm T}_k M_k = 1$ (the $(d - k) \times (d - k)$ matrix) at $k = 2, 3$. Substituting to $L$ we find
\begin{eqnarray}                                                           
L = \frac{A}{4 \pi G } \left [ - \frac{1}{ \gamma_{\rm F}} \tvn_2 (\varphi )^{\rm T} M_2( \theta )^{\rm T} M_1 (\chi )^{\rm T} \eta \frac{\d }{\d t} M_1 (\chi ) \tvn_1 (\theta ) \right. \nonumber \\ \left. + \tvn_3^{\rm T} M_3 (\varphi )^{\rm T} M_2( \theta )^{\rm T} M_1 (\chi )^{\rm T} \eta \frac{\d }{\d t} \vn_0 (\chi ) \right ].
\end{eqnarray}

\noindent Evidently, a part of the variables plays the role of canonical coordinates, then the other part plays the role of conjugate momenta. Therefore, in particular, the quantum state can be described by the functions of only a part of the variables. As the latter, it is most natural to consider $\chi$ and $\theta$. The variables $\chi$ describe noncompact rotations, therefore area being canonically conjugate to these does not lead to any discrete spectrum. As for the $\dot{\theta}$-terms in $L$, these are exactly
\begin{eqnarray}                                                           
& & L_{\dot{\theta}} = - \frac{A}{4 \pi G \gamma_{\rm F}} \tvn_2 (\varphi )^{\rm T} M_2( \theta )^{\rm T} \dottvn_1 (\theta ) \nonumber \\ & & = - \frac{A}{4 \pi G \gamma_{\rm F}} \left (B_1 \dot{\theta}_1 \sin \theta_{d-2} \dots \sin \theta_2 + \dots + B_{d-2} \dot{\theta}_{d-2} \right ).
\end{eqnarray}

\noindent The $B_1, \dots, B_{d-2}$ and $A$ can serve to parameterize the conjugate to $\theta_1, \dots, \theta_{d-2}$ momenta $p_n = \partial L / \partial \dot{\theta}_n$. In particular, we have for the area
\begin{equation}\label{AA}                                                 
\left ( \frac{A}{4 \pi G \gamma_{\rm F}} \right )^2 = \left (\frac{p_1}{\sin \theta_{d - 2} \dots \sin \theta_2} \right )^2 + \cdots +p^2_{d - 2}.
\end{equation}

\noindent In quantum theory, $p_n$ are substituted by the operators $-i\partial /\partial \theta_n$, and under appropriate product ordering the area operator squared (\ref{AA}) is nothing but minus angle part of the $(d - 1)$-dimensional Laplace operator. The spectrum of the latter is well-known (see, e.g., Ref. \cite{Vilen}). Its eigenfunctions are labeled by $d-2$ integers $(j, k_1, \dots, \pm k_{d-3})$ such that $j \geq k_1 \geq \dots \geq k_{d-3} \geq 0$. The eigenvalues $j(j + d - 3)$ depend on $j$ only. Then the number of eigenfunctions is
\begin{equation}                                                           
g(j) = \frac{(j + d - 4)!(2j + d - 3)}{j!(d - 3)!}
\end{equation}

\noindent for the $j$-th eigenvalue, that is, for the area
\begin{equation}                                                           
A \equiv 8 \pi l^2_p \gamma_{\rm F} a(j), ~~~ a(j) = \frac{1}{2} \sqrt{j (j + d - 3)}, ~~~ l^2_p = G.
\end{equation}

In the calculation of Ref \cite{Khr2}, $g(j)$ is the statistical weight of elementary areas $8 \pi l^2_p \gamma_{\rm F} a(j)$ with quantum number $j$. In this calculation, the requirement that the statistical formula for entropy would coincide with the Bekenstein-Hawking relation (which states that the entropy of the black hole is $(4 l^2_p)^{-1} A_{bh}$ where $A_{bh}$ is the horizon area) gives
\begin{equation}                                                           
\sum_j g(j) e^{-2 \pi \gamma_{\rm F} a(j)} = 1.
\end{equation}

\noindent This relation is an equation for $\gamma_{\rm F}$. Solving it for the genuine Faddeev's choice of dimensionality of the external space $d = 10$ we find
\begin{equation}                                                           
\gamma_{\rm F} = 0.39230246769....
\end{equation}

\noindent In principle, one can consider taking another $d$. For example, if the {\it global} embedding into the external Euclidean/Minkowskian space is considered, it may require as much as $d = 230$ dimensions \cite{230}. For this $d$ calculation gives
\begin{equation}                                                           
\gamma_{\rm F} = 0.359764688947....
\end{equation}

\noindent The dependence on $d$ is rather weak.

\section{ Conclusion}

The peculiarity of the situation compared with quantization of the angular momentum is that the area spectrum depends on some non-dynamical unconstrained (like the lapse-shift functions in GR, but not that in other respects) variables, here $b_\lambda$ having sense of certain rotation angles. It is not quite trivial that requiring consistency of the spectra of area operators constructed from different sets of conjugate pairs $p, q$ we could pick out the values of some $b_\lambda$s being $\pi$ (covering the flat spacetime as well) and get well-defined spectrum in the neighborhood of these. Although we used full nonlinear ansatz for the connection at the finite angles of rotation, in the end it came down to the analysis of the spectrum at $b_\lambda$s being in the neighborhood of $\pi$ or zero.

In LQG \cite{ABCK} $a(j) = \sqrt{j (j + 1)}$, $g(j) = 2j + 1$ where $j$ is a (half-)integer quantum number 1/2, 1, 3/2, ... , or $a(j) = \frac{1}{2} \sqrt{j (j + 2)}$, $g(j) = j + 1$ where now $j$ is an integer quantum number. This looks similar to our result for particular values of $d$, but not quite that. This similarity is due to the similarity of special orthogonal group transformation properties of the fundamental tetrad-connection variables in both cases. Another similarity is proportionality of area spectrum to $\gamma_{\rm F}$ or $\gamma$; in other words, the area spectrum being discrete is due to the parity-odd term in the action. In our approach, this is simply because the genuine (formed by both spacelike directions) area turns out to be canonically conjugate to the compact (Euclidean) part of the connection just due to the parity-odd term.

Thus, a concept of the discrete area spectrum exists in the Faddeev formulation of GR as well. One of the important properties providing this feature is the possibility of metric discontinuities and thus the possibility of the elementary areas being virtually independent.

\section*{Acknowledgments}

The present work was supported by the Ministry of Education and Science of the Russian Federation


\begin{thebibliography}{99}
\bibitem{Fad}
 L. D. Faddeev, New dynamical variables in Einstein's theory of gravity, {\it Theor. Math. Phys.} {\bf 166} 279 (2011); ({\it Preprint} arXiv:1003.2311[gr-qc]).
\bibitem{Chiu}
 Chiu Man Ho, Thomas W. Kephart, Djordje Minic, and Y. Jack Ng, Spacetime emergence and general covariance transmutation, {\it Mod. Phys. Lett.} A {\bf 28}, 1350005 (2013); ({\it Preprint} arXiv:1206.0085[hep-th]).
\bibitem{Regge'}
 T. Regge, General relativity theory without coordinates, {\it Nuovo Cimento} {\bf 19}, 558 (1961).
\bibitem{RegWil'}
 T. Regge and R. M. Williams, Discrete structures in gravity, {\it Journ. Math. Phys.} {\bf 41}, 3964 (2000); ({\it Preprint} arXiv:0012035[gr-qc]).
\bibitem{piecewiseflat=simplicial'}
 J. Cheeger, W. M\"{u}ller, and R. Shrader,  On the curvature of the piecewise flat spaces, {\it Commun. Math. Phys.} {\bf 92}, 405 (1984).
\bibitem{cdt}
 J. Ambjorn, A. Goerlich, J. Jurkiewicz, and R. Loll,  Nonperturbative Quantum Gravity, {\it Physics Reports} {\bf 519}, 127 (2012); ({\it Preprint} arXiv:1203.3591[hep-th]).
\bibitem{Ash}
 A. Ashtekar, C. Rovelli and L. Smolin, Weaving a classical geometry with quantum threads, {\it Phys. Rev. Lett.} {\bf 69}, 237-240 (1992); arXiv:hep-th/9203079.
\bibitem{Loll}
 R. Loll, Further results on geometric operators in quantum gravity, {\it Class. Quant. Grav.} {\bf 14}, 1725-1741 (1997); arXiv:gr-qc/9612068.
\bibitem{ABCK}
 A. Ashtekar, J. Baez, A. Corichi and K. Krasnov, Quantum Geometry and Black Hole Entropy, {\it Phys.Rev.Lett.} {\bf 80}, 904-907 (1998); arXiv:gr-qc/9710007.
\bibitem{LewDom}
 J. Lewandowski and M. Domagala, Black hole entropy from Quantum Geometry, {\it Class. Quantum Grav.} {\bf 21}, 5233-5244 (2004); arXiv:gr-qc/0407051.
\bibitem{Mei}
 K. Meissner, Black hole entropy in Loop Quantum Gravity, {\it Class. Quantum Grav.} {\bf 21}, 5245-5252 (2004); arXiv:gr-qc/0407052.
\bibitem{Barb}
 J. F. Barbero, Real Ashtekar Variables for Lorentzian Signature Space-times, {\it Phys.
 Rev. D} {\bf 51}, 5507-5510 (1995); arXiv:gr-qc/9410014.
\bibitem{Imm}
 G. Immirzi, Quantum Gravity and Regge Calculus, {\it Nucl. Phys. Proc. Suppl.} {\bf 57}, 65-72 (1997); arXiv:gr-qc/9701052.
\bibitem{Holst}
 S. Holst, Barbero's Hamiltonian Derived from a Generalized Hilbert-Palatini Action,
 {\it Phys. Rev. D} {\bf 53}, 5966-5969 (1996); arXiv:gr-qc/9511026.
\bibitem{Fat}
 L. Fatibene, M. Francaviglia and C. Rovelli, Spacetime Lagrangian For\-mu\-la\-ti\-on of Barbero-Immirzi Gravity, {\it Class. Quantum Grav.} {\bf 24}, 4207-4218 (2007); arXiv:0706.1899.
\bibitem{Bek2}
 J. D. Bekenstein, Universal upper bound to entropy-to-energy ratio for bounded systems, {\it Phys. Rev. D} {\bf 23}, 287-298 (1981).
\bibitem{Tho}
 G. 't Hooft, Dimensional Reduction in Quantum Gravity, in {\it Salam Festschrift} (Singapore, 1993); arXiv:gr-qc/9310026.
\bibitem{Sus}
 L. Susskind, The World as a Hologram, {\it J. Math. Phys.} {\bf 36}, 6377-6396 (1995); arXiv:hep-th/9409089.
\bibitem{Khr1}
 I. B. Khriplovich and R. V. Korkin, How Is the Maximum Entropy of a Quantized Surface Related to Its Area?,
 {\it J. Exp. Theor. Phys.} {\bf 95}, 1-4 (2002); {\it Zh. Eksp. Teor. Fiz.} {\bf 95}, 5-9 (2002); arXiv:gr-qc/0112074.
\bibitem{Glo}
 A. Ghosh and P. Mitra, An improved estimate of black hole entropy in the quantum geometry approach, {\it Phys. Lett. B} {\bf 616}, 114-117 (2005); arXiv:gr-qc/0411035.
\bibitem{Khr2}
 I.B. Khriplovich, Quantized Black Holes, Their Spectrum and Radiation,
 {\it Phys. Atom. Nucl.} {\bf 71}, 671-680 (2008); arXiv:gr-qc/0506082.
\bibitem{Cor}
 A. Corichi, J. Diaz-Polo and E. Fernandez-Borja, Quantum geometry and microscopic black hole entropy, {\it Class. Quant. Grav.} {\bf 24}, 243-251 (2007); arXiv:gr-qc/0605014.
\bibitem{II}
 V. M. Khatsymovsky, Some minisuperspace model for the Faddeev formulation of gravity, {\it Mod. Phys. Lett.} A {\bf 29}, 1450141 (2014); {\it Preprint} arXiv:1408.6375[gr-qc].
\bibitem{I}
 V. M. Khatsymovsky, First order representation of the Faddeev formulation of gravity,  {\it Class.Quant.Grav.} {\bf 30} 095006 (2013); {\it Preprint} [arXiv:1201.0806 [gr-qc].
\bibitem{III}
 V. M. Khatsymovsky, First order minisuperspace model for the Faddeev for\-mu\-la\-ti\-on of gravity, {\it Mod. Phys. Lett.} A {\bf 30}, 1550172 (2015); {\it Preprint} arXiv: 1508.07573[gr-qc].
\bibitem{Kha}
 V. M. Khatsymovsky, On area spectrum in the Faddeev gravity, {\it Preprint} arXiv: 1206.5509[gr-qc].
\bibitem{Vilen}
 N. Ya. Vilenkin, {\it Special Functions and the Theory of Group
 Representations}, Translations of Mathematical Monographs, Vol. 22 (Amer.
 Math.  Soc., Providence, Rhode Island, 1968).
\bibitem{230}
 J.F. Nash, The imbedding problem for Riemannian manifolds,
 {\it Ann. Math.} {\bf 63}, 20-63 (1956).
\end{thebibliography}
\end{document}